\documentclass[prl,twocolumn,epsf,psfig]{revtex4}
\usepackage{graphicx}
\DeclareGraphicsExtensions{.pdf,.png,.gif,.jpg}
\usepackage{float}
\usepackage{subfig}

\begin{document}

\title{Evidence for the breakdown of momentum independent many-body t-matrix approximation in the normal phase of Bosons}

\author{Theja N. De Silva}
\affiliation{Department of Chemistry and Physics,
Georgia Regents University, Augusta, GA 30912, USA.}
\begin{abstract}
We revisit the momentum independent many-body t-matrix approach for boson systems developed  by Shi and Griffin~\cite{sg} and Bijlsma and Stoof~\cite{bs}. Despite its popularity, simplicity, and expected advantage of being its applicability to both normal and superfluid phases, we find that the theory breaks down in the normal phase of bosons. We conjecture that this failure is due to neglecting of momentum dependence on the t-matrix.
\end{abstract}

\maketitle

The impressive experimental realization of weakly interacting Bose-Einstein condensation (BEC) in dilute alkali gases~\cite{becE1, becE2, becE3} has renewed the theoretical and experimental interest in studies of Bose systems. The BEC of alkali gases were identified by the existence of a sharp peak in the momentum distribution below a certain critical temperature. Unlike the intensively studied strongly interacting liquid $^4$He system, the momentum distribution of cold atomic systems is easily accessible due to the inhomogeneity of the trapped atoms. While, the focus of the first generation of cold gas experiments has been on BEC of weakly interacting bosons, the center of research has now shifted towards the study of both normal and superfluid phases in strongly interacting bosons. The main advances in this direction have been achieved by increasing the
interaction through controlling the two-body scattering length ($a$) or exposing the atoms to optical lattices~\cite{sl1}.

Impressive theoretical efforts are currently devoted to study the effect of interactions on Bose-Einstein condensation~\cite{th1, th2, th3}. Until recently, there was no consensus on the sign of the shift of critical temperature in the presence of strong interactions. Now it is believed that the shift can be described as $\Delta T_c = 1.3 a n^{1/3} T_c^0$, where $n$ is the density and $T_c^0$ is the critical temperature of an ideal Bose gas~\cite{ct1,ct2}. Further, it has been shown that none of the mean field theories correctly predict the expected second-order BEC-normal phase transition~\cite{lovr}. These mean field theories include, Hartree-Fock (HF)~\cite{th3,hf}, Popov~\cite{pop}, Yukalov-Yukalova (YY)~\cite{yy}, and t-matrix approximations~\cite{sg, bs}. Recently, a self-consistent mean field theory was proposed as the \emph{sole} mean-field theory that explains the correct second-order transition~\cite{tmn}. All theories mentioned above have been applied to study either the properties of the BEC phase or the phase transition. Little attention has been paid to the normal phase.

One of the most popular and intensively applied theories in literature was proposed by Bijlsma and Stoof~\cite{bs} and Shi and Griffin~\cite{sg}. This t-matrix approximation (TM) was developed based on Popov theory. The authors in references~\cite{sg} and \cite{bs} however go beyond the contact interaction and include the many-body effects by taking into account higher order scattering. Then neglecting the momentum dependence on the t-matrix, these authors have derived a simple analytical expression for density in both normal and superfluid phases. The expected advantage of TM theory is that it is applicable for both superfluid BEC and normal phases. In this opinion article, we show that this t-matrix approximation based theory breaks down in the normal phase and the expected advantage and simplicity of the theory is no longer valid.

\textbf{\emph{TM theory:}} First we summarize the TM approach for the normal phase of bosons, following the reference~\cite{sg}. The density of the normal bosons is $n = g_{3/2}[\beta \Delta]/\lambda^3$, where $\lambda = [2 \pi \hbar^2/(mk_BT)]^{1/2}$ is the thermal deBroglie wavelength and $g_{n}(x)$ is the well known Bose integral. The parameter $\Delta$ is related to the density $n$, chemical potential $\mu$, many-body t-matrix $T = U/(1+ \alpha U$), and an additional parameter $\alpha$ through the expression $\mu = \Delta+ (2nU)/(1+\alpha U)$, where

\begin{eqnarray}
\alpha = \sum_k\biggr(\frac{1}{2E_k}
coth (\beta E_k/2)-\frac{1}{2\epsilon_k}\biggr).
\end{eqnarray}

\begin{figure}
\includegraphics[width=\columnwidth]{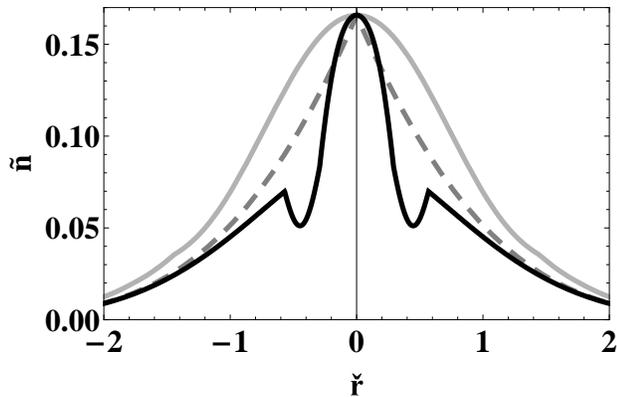}
\caption{The scaled density profiles of a harmonically trapped Bose gas at the onset of BEC. While the center of the trap ($\check{r} =0$) is set to be at BEC transition, the entire tail of the density profile is in normal state. The interaction strength and temperature are fixed to be $U =\hbar \omega$ and $k_B T =\hbar \omega$. The dashed-line represents a non-interacting Bose gas, the black and gray lines represent t-matrix and Hartree-Fock theories respectively. The density $\tilde{n} = n l^3$ and the radial coordinate $\check{r} = r/l$ are scaled with the oscillator length $l = \sqrt{\hbar/m\omega}$.}\label{SG}
\end{figure}

\noindent Here $U = 4\pi \hbar^2a/m$ is the free-space scattering amplitude, $\hbar$ is the plank constant, $k_B$ is the Boltzmann constant, $\beta = 1/(k_BT)$ is the inverse temperature, and $m$ is the mass of a boson. The quasi particle energy is given by $E_k = \epsilon_k-\Delta$ with $\epsilon_k = \hbar^2 k^2/2m$. At the normal-BEC transition, $\Delta \rightarrow 0$, while $\alpha \rightarrow \infty$ yielding a chemical potential $\mu_c =0$ as in the case of an ideal Bose gas. In order to demonstrate the problematic behavior of the theory, we solve these set of equations for the density profile of a harmonically trapped Bose gas. Using local density approximation, replacing $\mu \rightarrow \mu_0 - m\omega^2r^2/2$, the density profile of the trapped gas is shown in FIG.\ref{SG} as a function of spatial coordinate $r$. Here $\mu_0$ is the chemical potential at the center of the trap and $\omega$ is the trapping frequency. The density profile is shown at the onset of the BEC. In other words, the trap center $r =0$ is set to be at the BEC where we set central chemical potential to be $\mu_0 = \mu_c$.

\textbf{\emph{HF theory:}} As a comparison, we show HF density profile for the same parameters in gray color in FIG.\ref{SG}. HF approach is a self-consistent approach that simplifies the N-particle interacting states into effectively one-particle non-interacting states whose energy spectrum depends self-consistently on both density and the interaction. The HF normal density is given by $n = g_{3/2}[\beta (\mu-2nU)]/\lambda^3$. At the BEC transition, the chemical potential is $\mu_c = 2Ug_{3/2}[0]/\lambda^3$. It is worth pointing out that each of the mean field theories, HF, Popov, and YY are equivalent in the normal phase but they differ in the BEC phase. Notice that the critical density at the transition is the same as that of the TM approach and the non-interacting Bose gases, but the critical chemical potential is different. The non-interacting density profile at the onset of BEC is also shown as a dashed line.

The homogenous density in the normal phase is expected to be a monotonic function of chemical potential. Likewise, the density profile in a trap is expected to be a monotonically decreasing function of the spatial coordinate $r$. However, as opposed to the HF theory, the TM approach gives non-monotonic density variation. The density at the vicinity of BEC (density of the gas at the trap center is at criticality) however, agrees within the two approaches. The results are shown for representative values of interaction and temperature. The qualitative behavior is very similar for all finite interactions and temperatures. Even though this TM approach is widely used in literature, this problematic behavior has not been previously reported. This may be due to the fact that most studies focus on the BEC phase (where the density is monotonic as expected), but not on the normal phase. Though we use the local density approximation for a harmonically trapped boson system to demonstrate this ill behavior, the same behavior exists as a function of chemical potential. In other words, the density is not a monotonically increasing function of chemical potential. Therefore, it is not the local density approximation, but the approximation made to the t-matrix approach that must be responsible for this problematic behavior.

In conclusion, we have revisited the normal state bosons using a momentum independent t-matrix approach and discovered that the expected validity of the theory breaks down in the normal phase. Although this theory has been applied in numerous previous studies, this ill behavior has not been previously reported as to the best of our knowledge. We anticipate that the theory can be recovered by the inclusion of momentum dependence on the t-matrix. This will be a non trivial task due to the infrared divergences and including these will destroy the simplicity of the theory.

\textbf{\emph{Acknowledgements: }} We are grateful to Joseph Newton for critical comments on the manuscript.

\end{document}